# Multiple Sclerosis Lesions Segmentation using Attention-Based CNNs in FLAIR Images

Mehdi SadeghiBakhi, HamidReza Pourreza, Hamidreza Mahyar

**Objective:** Multiple Sclerosis (MS) is an autoimmune, and demyelinating disease that leads to lesions in the central nervous system. This disease can be tracked and diagnosed using Magnetic Resonance Imaging (MRI). Up to now a multitude of multimodality automatic biomedical approaches is used to segment lesions which are not beneficial for patients in terms of cost, time, and usability. The authors of the present paper propose a method employing just one modality (FLAIR image) to segment MS lesions accurately.
**Methods:** A patch-based Convolutional Neural Network (CNN) is designed, inspired by 3D-ResNet and spatial-channel attention module, to segment MS lesions. The proposed method consists of three stages: (1) the contrast-limited adaptive histogram equalization (CLAHE) is applied to the original images and concatenated to the extracted edges in order to create 4D images; (2) the patches of size $80 \times 80 \times 80 \times 2$ are randomly selected from the 4D images; and (3) the extracted patches are passed into an attention-based CNN which is used to segment the lesions. Finally, the proposed method was compared to previous studies of the same dataset.
**Results:** The current study evaluates the model, with a test set of ISIB challenge data. Experimental results illustrate that the proposed approach significantly surpasses existing methods in terms of Dice similarity and Absolute Volume Difference while the proposed method use just one modality (FLAIR) to segment the lesions.
**Conclusions:** The authors have introduced an automated approach to segment the lesions which is based on, at most, two modalities as an input. The proposed architecture is composed of convolution, deconvolution, and an SCA-VoxRes module as an attention module. The results show, the proposed method outperforms well compare to other methods.

## I. INTRODUCTION[1]

MULTIPLE sclerosis is an autoimmune, chronic, and demyelinating disease of axons that causes lesions in the brain's white matter (WM) tissues [1]. Varying from patient to patient, the most common symptoms of MS are weakness, balance issues, depression, fatigue, or visual impairment. As the most prominent visualization method for medical imaging nowadays, Magnetic Resonance Imaging (MRI) can visualize and diagnose this kind of disease [2]. Accurate segmentation of MS lesions in MR images is one of the most critical tasks in figuring out and describing the progression of the disease [3]. To do so, manual and automated segmentation methods are commonly employed to estimate and segment the total number of lesions and total lesion volume. Although manual segmentation is considered a desirable standard method [4], it poses challenges in the description of 3-dimensional (3D) MRI information. This method is found to be time-consuming, annoying, and prone to intra- and inter-observer variability. These challenges motivate Deep Learning (DL) and Machine Learning (ML) researchers to propose and develop a fast and accurate approach for the segmentation of MS lesions in MRI [5].

Supervised machine learning algorithms are one group of automated methods that can acquire knowledge from previously labeled training data and present high efficiency in MS lesion segmentation [6]. Generally, traditional supervised machine learning approaches are dependent on hand-crafted or low-level features. So far, plenty of supervised techniques for MS lesion segmentation have been proposed, such as decision random forests [7] [8], ensemble methods [9], non-local means [10], k-nearest neighbors, [11] [12] and combined inference from patient and healthy populations [13]. Another group of automated methods is unsupervised which extracts patterns from unlabeled data. Therefore, a great number of unsupervised methods have also been introduced, which are based on thresholding methods with post-processing to remove FP and FN pixels [14] [15] or probabilistic models [16][17].

Since 1988, deep learning methods, especially convolutional neural networks (CNNs), have significantly increased performance in biomedical image analysis [18]. They require fewer manual features than standard supervised ML algorithms and can learn by themselves how to extract features directly from data during the training procedure [19]. Deep learning-based approaches provide state-of-the-art results for different problems such as computer vision semantic segmentation [20], as well as natural language processing [21]. They have also gained popularity in studying biomedical problems, such as cell classification [22], retinal blood vessel extraction [23], MS lesions [24], brain tumors [25], neuronal structures [26], and brain tissue segmentation [27].

A great number of CNN-based algorithms for biomedical image segmentation have been proposed. These approaches can be divided into image-based and patch-based methods. Details of these approaches are discussed in the following paragraph.

Image-based methods, extract features from the whole image as global structure information [28] [29] and are categorized into 2D-based [28] [30] and 3D-based [29] [31] segmentation.

In 3D-based methods, features are first extracted from an original 3D image by employing CNNs through 3D filters. The model then segments each pixel or voxel to the lesion or non-lesion. The high chances of overfitting are one of the disadvantages of the 3D-based method, which usually fits a large number of parameters when the dataset is small. Regrettably, this is a common occurrence in biomedical applications [29] [31].

In 2D-based methods, the 3D image is first divided into 2D slices and, then each slice is eventually segmented. Finally, to reconstruct the 3D prediction, all 2D predictions are concatenated together. When compared to 3D-based methods, 2D-based approaches are not as accurate due to missing part of the contextual information. However, they have fewer parameters for each layer, which can lower the risk of overfitting in small datasets [30].

Patch-based methods use two different strategies for lesion segmentation. Utilizing a moving window, the first strategy creates a local representation for each pixel/voxel. Then, a CNN is trained to use all the extracted patches to identify the central pixel/voxel of each patch as either a lesion or a non-lesion. Considering the repetitive computations on the overlapping features of the sliding window, it is worth mentioning there is a long training time. Faster than the previous strategy, the second approach, randomly extracts patches from all over the 3D image. The extracted patches are then used to train the CNN-based model. Ultimately, the trained model segments each pixel/voxel of the patch as either a lesion or a non-lesion [32].

The rest of the current paper is organized into five sections and a reference section. Section II reviews works related to the proposed method. Section III explains the dataset used to train and evaluate the proposed method. Section IV clarifies the methodology. Section V explains the evaluation metrics and compares the results of the test dataset against the findings of other methods. Finally, a discussion and summary are provided in Section VI.

Mehdi SadeghiBakhi was in Department of Computer Engineering, Faculty of Engineering, Ferdowsi University of Mashhad, Mashhad, Iran (e-mail: Mehdi.sadeghibakhi@mail.um.ac.ir)

HamidReza Pourreza is a computer engineering professor in Department of Computer Engineering, Faculty of Engineering, Ferdowsi University of Mashhad, Mashhad, Iran (e-mail: Hpourreza@um.ac.ir)

HamidReza Mahyar is a professor of W Booth School of Engineering Practice and Technology, Faculty of Engineering, McMaster University, Canada (e-mail: mahyarh@mcmaster.ca)
.

*A. Motivation*

For automatic MS lesion segmentation, the present paper proposes a new method based on deep learning which consists of two 3D convolutional network branches. The current study randomly extracts some large 3D patches to prevent overfitting due to the small number of data, and the lack of global structure information. Then a deep neural network is designed inspired by 3D version of ResNet [33], which consists of convolution and deconvolution layers, channel-based attention, and special-based attention.

The designed network can be employed in each path and each branch is assigned to a particular MRI modality so that the maximum benefits of each modality can be individually utilized. As a result, the present study has introduced a network that can have at most two different modalities (FLAIR and T1) However, for the first time, stunning results were achieved with just one modality (FLAR), when compared to other methods. In summary, the main contributions in this paper are:
- Introduction of a 3D patch-based method to prevent overfitting and a lack of global structure information and to combine the two modalities in order to take advantage of brain tissue information.
- Proposal of a deep learning-based approach with one modality (FLAIR) to segment the MS lesion which, for the first time, due to, the difficulty of providing variant modalities in terms of cost and time in the clinical situation.
- Proposal of a deep learning-based approach with one modality (FLAIR) to segment the MS lesion which, for the first time, due to, the difficulty of providing variant modalities in terms of cost and time in the clinical situation.
- Design of a 4D channel-wise and spatial attention module inspired by channel-based [34] attention and special-based attention [35].

*B. Related works*

As already stated, a wide range of methods for MS lesion segmentation has been proposed. Recently, convolution neural network strategies have reported outstanding performance in medical image processing, especially in MS lesion segmentation. Hence, this discussion of related works is related to CNN-based methods.

It should be noted that some approaches are patch-based. A useful illustration is a study by, Ghafoorian et al. [36] in 2015 which proposed a 2D CNN-based model that, increases the number of training samples and avoids overfitting during the training process. Similarly, Birenbaum et al. introduced multiple 2D patch-based CNNs that use in parallel the benefits of the common information within longitudinal data [37]. Roy et al. developed a two-path CNN based on a 2D-patch which employs different MRI modalities in each path as an input and finally concatenates the output of each path to create the final prediction [32]. Afzal et al. proposed a system that includes two 2D patch-wise CNNs which accurately segment lesions. Their first CNN network creates a lesion map while the second enhances efficiency by reducing the number of false positives [38]. Shachor et al. introduced a multi-view structure based on 2D patches. Each view of the 2D patches is passed to the model as an input and the patches output is fused to create the final prediction [39].

In addition, there are some approaches to segment lesions based on 3D patches. Vaidya and Suthirth presented a 3D patch-based CNN for segmenting lesions, after which a WM mask is applied to the output prediction to reduce the False Positive (FP) rate and attain high performance [40]. Valverde et al. [24] proposed a cascaded 3D CNN approach whose first model is trained with extracted 3D patches and its second model is then used to reduce the FP of the first model. In addition, Valverde et al. developed a model to analyze the effect of intensity domain adaptation on CNN-based models [41]. Simon et al. introduced a method based on multi-dimensional gated recurrent units and used 3D patches to train the model [42]. Salem et al. [43] presented a CNN-based model to create synthesis lesions in MR images as a way to deal with one of the biggest challenges in medical image processing, that is, a small number of data. They reported acceptable results even though only one image is used as a dataset to train Valverde's proposed model [24]. Hashemi et al. [44] proposed a 3D patch-based CNN method which employs the idea of a densely connected network. They also introduced a new loss function to deal with imbalanced data [44].

Furthermore, some approaches have used the whole image as an input. Brosch et al. proposed a whole brain-based segmentation method utilizing 3D CNN which takes advantage of some shortcut connections between layers so as to extract the low- and high-level features from the shallowest to the deepest layers. By doing so, the model learns information and features about the locations and structure of MS lesions [29]. Chuan hu et al. proposed an attention context U-NET based on 3D images [31]. Aslani et al. designed a deep 2D encoder-decoder CNN for the segmentation of MS lesions [45]. Another paper of theirs introduced, a method based on a 2D CNN for slice-by-slice segmentation of lesions in 3D data. Lesions are separately segmented on each slice, and then each segmented slice is concatenated to create a 3D lesion mask [30].

Although all of the proposed patch-based techniques perform well in terms of segmentation, they all lack global structural details. Simply put, the segmentation process does not take into account the brain's overall structure or the exact location of lesions. In contrast, whole brain-based segmentation methods need a large number of data to train the model, which is regrettably a commonplace requirement in biomedical applications that is difficult to meet.

## II. METHODS AND PROCEDURES

*A. Material*

To evaluate and compare the proposed method's performance to that of other state-of-the-art approaches, the present paper employs the ISBI[2] 2015 Longitudinal MS Lesion Segmentation Challenge dataset, which is publicly

---

[2] International Symposium on Biomedical Imaging

available on the challenge website. Further details are provided in the upcoming section.

*ISBI 2015 Longitudinal MS Lesion Segmentation Challenge:*

The ISBI dataset includes 19 subjects which are divided into a test set with 14 subjects and a training set with five subjects. For each subject, there are varieties of time-points in the range of 4 to 6, for each of which T1w, T2w, PDW, and FLAIR image modalities are prepared. The size of each image is 182 × 256 × 182 and the voxel resolution is 1 millimeter. Two different raters (R1 and R2) manually segment the images, so the data set has two ground truth lesion masks. The ground truth is publicly available for training images, but not for test images. However, the proposed method's performance over the test set is evaluated by submitting binary masks to the challenge website [46].

### B. Method

The process of constructing the model consists of three stages: first, the preprocessing data, second, the patch extraction from images, and last, lesion prediction using the trained model. The following discusses all processes of the proposed method in detail.

#### 1) Data Preprocessing

The present study utilizes the preprocessed version of the images available on the challenge website. Preprocessing algorithms that are already applied to images of the dataset are skull-stripped by the Brain Extraction Tool (BET) [47] and N3 intensity normalization [48] and rigidly registered to the MNIICBM152 template [49].

The current work applies more preprocessing algorithms to enhance the local contrast of the images and to avoid the distorting differences in the ranges of values. Contrast-Limited Adaptive Histogram Equalization (CLAHE) [50] is applied to achieve this objective. Then, the edges extracted by the Laplacian detectors are concatenated to enhance the image and the 4D data is created. Finally, before passing the data into the network, the intensities of each image are normalized with a zero mean and unit variance. Figure 1 presents an illustrated example of one slice of an extracted edge and the enhanced image.

#### 2) Patch extraction

Extraction of patches from the images begins after data preparation. The input of the model is a bunch of patches of images. Approximately 60% of the selected patches, which are 80 × 80 × 80 × 2 in size, are centralized on the lesion voxel. The rest of the patches are centered on the non-lesion voxel. Figure 2 depicts some of the extracted patches.

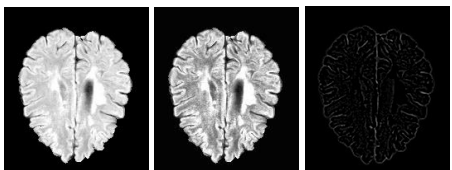

Figure 1: Impact of a preprocessing algorithm on the fourth screening of the sixth training sample. From left to right, a slice of the 3D original image, the preprocessed version of the image, and the extracted edge by the Laplacian filters.

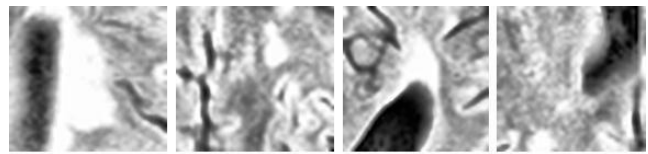

Figure 2: Samples of extracted patches from the dataset images. From left to right, the first and third patches are centered on the lesion voxel, while the others are centered on the non-lesion voxel.

#### 3) Network Architecture

This section presents the proposed architecture outline. Inspired by the spatial attention [35] and channel attention [34] strategies, the present study integrates an adapted version of these two attentions into 3D ResNet architecture in order to capture a better contextual representation of the image. Subsequently, decision-level fusion [51] is employed to learn the complementary information independently from the different modalities.

##### a) Base model

Inspired by the 3D ResNet, the architecture of a base model for MS lesion segmentation is illustrated in Figure 3. The model consists of convolutional, deconvolutional, batch normalization (BN), and rectified linear unit (ReLU) layers, as well as six stacked residual modules (i.e., SCA-VoxRes modules) with a total of 25 volumetric convolutional/deconvolutional layers. As shown in Figure 4, each SCA-VoxRes module includes two convolutions, two BN/ReLU layers, and an SCA module. In this module, the transformed feature and input feature are added together by the skip connection. With this connection, information can be propagated directly to the forward and backward passes. In addition, the SCA module includes spatial and channel-wise attention which will be explained upcoming sections.

It should be pointed out that filters and operations are implemented in a 3D shape to learn and extract stronger volumetric feature representation. Due to their computation efficiency and representation capability, the small kernels (i.e $3 \times 3 \times 3$) are employed in the convolutional layers. To reduce the resolution of the input image and features, three convolutional layers are employed with a stride of 2. As a result, a large receptive field network is obtained that can extract more contextual information to improve discrimination capability. To overcome the internal covariance shift in the training process and improve network performance, four BN layers are inserted into the network. In this network, rectified linear units are used as the activation function for nonlinear transformation. Lastly, the extracted features in the 3rd, 5th, 9th, and 15th layers are deconvolved and concatenated together to use as an input of the segmentation step.

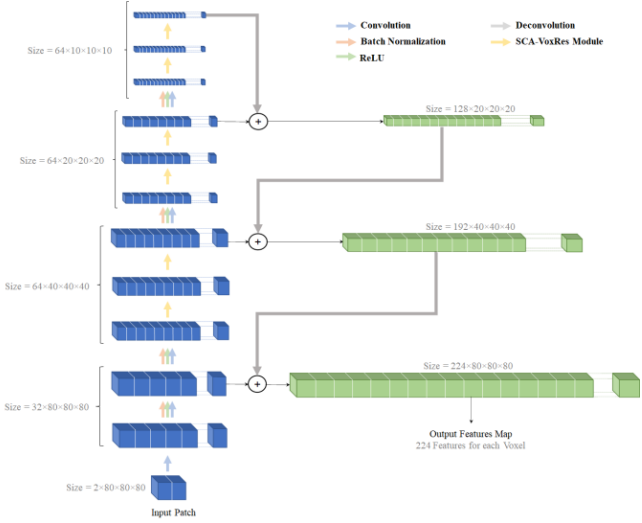

Figure 3: Overview of the proposed Base Model (CNN-based architecture to extract features for lesion segmentation) as described in Section II. Data is presented in an input by 4D patches (The first dimension is the number of input channels, and then are the depth, height, and width respectively) and the model generates the feature maps which have 224 features for each voxel of input patch. Finally, the output of the Base Model is used as input to the segmentation layer, which is responsible for the segmentation of each voxel to lesion or non-lesion.

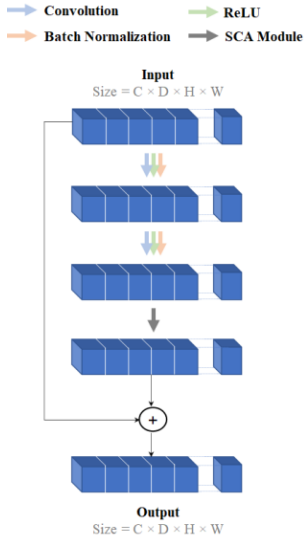

Figure 4: SCA-VoxRes Module. The input should be in the size of (C, D, H, W) where C is the number of channels, D is the depth, H and W are heigh and width respectively.

*b)    Spatial and Channel-wise Attention*

Diagnosis of the brain's white matter lesions is very challenging due to the presence of similar pixels/voxels in brain tissue and the resulting intensity. Furthermore, in CNN-based approaches, convolution operators represent a local receptive field. Consequently, the features extracted from pixels with the same label may somewhat differ and these differences cause intra-class inconsistency in addition to affecting the accuracy of the model. For this reason, an attention module is developed, as illustrated in Figure 5, which is based on a combination of spatial and channel-wise attention for exploring global contextual information.

***Channel-wise Attention:*** The objective of channel-wise attention is to enhance the network feature representation capability and emphasize interdependent feature map-specific semantics by exploiting the interdependencies between the channels of the feature maps. The structure of channel-wise attention is illustrated in Figure 5. First, the channel-wise attention map, $X \in \mathbb{R}^{1 \times 1 \times 1 \times C}$, is directly calculated by an average pooling and two fully connected layers from the original features, $F \in \mathbb{R}^{D \times H \times W \times C}$. Then, the channel-wise attention map is multiplied by the input feature to scale each channel of the feature map.

***Spatial Attention:*** The extraction of discriminatory representation is essential for the segmentation of WM lesions and this can be achieved by capturing long-range contextual information. Thus, a spatial attention mechanism can encode a global representation with a wider field of view into local features. As illustrated in Figure .5, the input feature map, $F \in \mathbb{R}^{D \times H \times W \times C}$, is fed into the three one-by-one convolution layers to generate three new feature maps, called $A$, $B$, $C$, while they are the same size, $\mathbb{R}^{D \times H \times W \times C}$. Next, $A$ and $B$ are reshaped to $\mathbb{R}^{N \times C}$, where $N = D \times H \times W$ is the number of voxels. Later, the spatial attention map, $E \in \mathbb{R}^{N \times N}$, is calculated by the matrix multiplication of $B$ and the transpose of $A$, followed by the application of a SoftMax function to the result. In the next step, the feature map, $C \in \mathbb{R}^{D \times H \times W \times C}$, is reshaped to the $\mathbb{R}^{N \times C}$ matrix. Then, matrix multiplication is performed between $C \in \mathbb{R}^{N \times C}$ and $E \in \mathbb{R}^{N \times N}$ and this reshapes the result to $\mathbb{R}^{D \times H \times W \times C}$. Finally, the output of spatial attention is calculated by Equation 1.

$$O_j = \omega \sum_{i=1}^{n} \left( E_{ij} C_i \right) + F_j \quad (1)$$

where $\omega$ refers to a learnable parameter. The output of spatial attention can be obtained from Equation 1 as each voxel is a weighted sum of the features.

*4) Two-path Architecture*

In medical image processing applications, datasets provide different imaging modalities for robustly analyzing various tissue structures. For example, in the ISBI dataset, four modalities are available, such as T1w, T2w, PDW, and FLAIR. The most significant reason for providing multi-modality images is because the information of the modalities complements each other. Therefore, utilizing the basic model, the current study designed a decision-level fusion architecture with two individual paths. Figure 6 presents this architecture. First, two modalities, T1w, and FLAIR are individually processed and two different feature maps are extracted. Second, the extracted feature maps from both paths are concatenated together and then used as an input of classification. This strategy allows for individual feature learning for each modality before the aggregation of the feature maps.

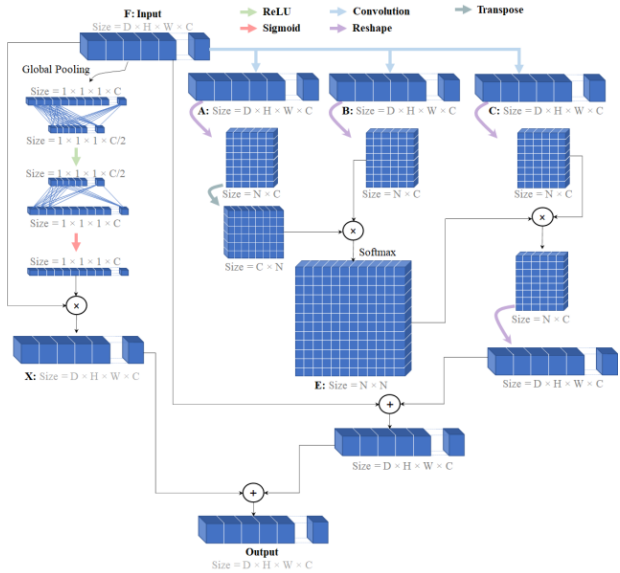

Figure 5: Spatial Channel Attention (SCA) module based on 4D input. D, H, W, C are depth, height, width and channel. As well as N is equal to $D \times H \times W$.

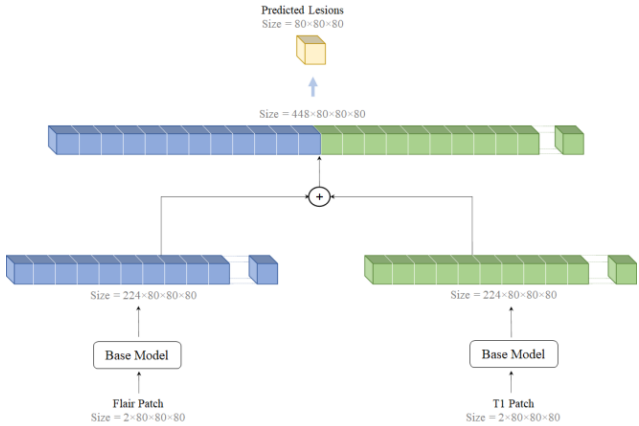

Figure 6: The architecture of the Two-path Network integrates the feature maps, which are prepared by the Base Model, and segments them using a one-by-one convolution layer.

### C. Loss Function

One of the remarkable problems in medical image processing algorithms is dealing with unbalanced data. Since 3D images have fewer lesion voxels than non-lesion voxels, an imbalance in data is faced. Therefore, if the used loss function is unable to deal with this problem, the model converges to the minority class. In other words, samples in a class with a small number (minority class) cannot be classified accurately by using the model. To handle the imbalanced data, the proposed method uses a combination of Tversky [44] and Focal [52] loss functions. As illustrated in Equation 2, the Tversky loss function allows the assignment of different weights to False Negative (FN) and False Positive (FP) in order to improve the recall rate.

$$TverskyLoss = \sum_c 1 - \frac{\sum_{i=1}^{N} p_{ic} g_{ic} + \varepsilon}{\sum_{i=1}^{N} p_{ic} g_{ic} + \alpha \sum_{i=1}^{N} p_{\bar{i}c} g_{ic} + \beta \sum_{i=1}^{N} p_{ic} g_{\bar{i}c} + \varepsilon} \quad (2)$$

where $p_{ic}$ is the probability that pixel $i$ is of lesion class $c$, and $p_{\bar{i}c}$ is the probability that pixel $i$ is of non-lesion class $\bar{c}$, and the same can be said for $g_{ic}$ and $g_{\bar{i}c}$ which are related to the grand truth. In addition, hyperparameters, $\alpha$ and $\beta$, can be tuned by assigning a number in the [0, 1] range.

The Tversky loss function limitation is the low convergence speed due to the segmentation of small ROIs. This does not contribute significantly to losses. However, to overcome this problem, the current work utilizes the Focal Tversky loss function, which has a parameter, $\gamma$, to control the segmentation of small ROIs. Equation 3 defines the Focal Tversky Loss.

$$FocalTverskyLoss = \sum_c 1 - \left( \frac{\sum_{i=1}^{N} p_{ic} g_{ic} + \varepsilon}{\sum_{i=1}^{N} p_{ic} g_{ic} + \alpha \sum_{i=1}^{N} p_{\bar{i}c} g_{ic} + \beta \sum_{i=1}^{N} p_{ic} g_{\bar{i}c} + \varepsilon} \right)^\gamma \quad (3)$$

### D. Train and Test details

At the beginning of the training procedure, the data should be split to make the train, test, and validation dataset. The ISBI dataset includes two sets of images. Consisting of 21 images from five subjects, the first set of images with available ground truth is randomly divided into validation and training sets. Sixty images are considered as a train set and five images as a validation set. In addition, the second set of images is used as a test set.

One of the main challenges for gradient-based optimization methods is choosing an optimal starting point for the learning rate. Classical methods employ a fixed number for the learning rate in all stages of training, however, it is better to adjust the learning rate during the training procedure by reducing it according to a predefined schedule function. In this paper, exponential decay is used as the schedule function. In our experiments, the Adam [53] optimizer outperforms performed better than other optimizers, such as AdaGrad [54], AdaDelta [55], and RMSprop [56]. So, Adam is selected as an optimizer.

In order to test the model, the input image is divided into $80 \times 80 \times 80$ image patches. The model predicts the label for each part and, in the end, all predictions are integrated as the label of the given image.

### E. Implementation Details

The proposed method is implemented in the Python programming language with the Pytorch framework. The experiments are performed on Google Colaboratory[3] with 12 GB RAM. The two-path network is trained end-to-end and, to do so, 4D patches are randomly extracted from the 4D data as described in Section II. Then, the Focal Tversky loss function deals with the imbalanced data problem, as explained in Section II, with $\alpha = 0.7$, $\beta = 0.3$, $\gamma = 4/3$.

---
[3] colab.research.google.com

In addition, to optimize the network parameters and find the best model, the current work utilizes validation data and saves the model with the best validation data results. The model is trained in 100 epochs with the Adam optimizer at an initial learning rate of 0.0001. In each epoch, the learning rate is reduced by the exponential decay function. According to Figure 7, the best performance obtained is at epoch 80. The training computation time of 100 epochs is approximately 12 hours.

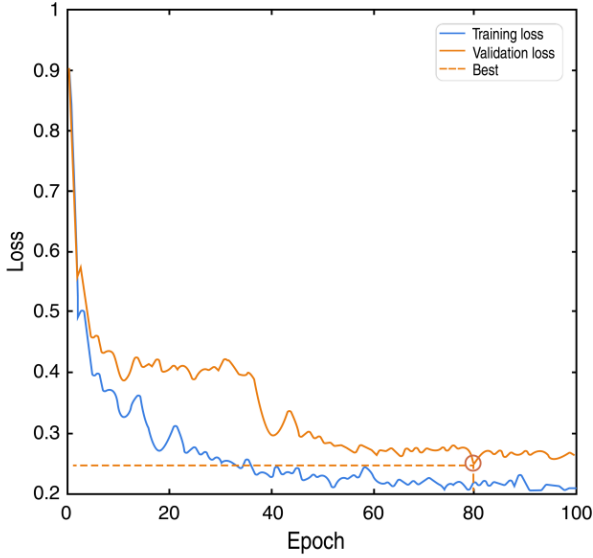

Figure 7: The decrease of loss function during the training step on the ISBI dataset. Due to overfitting, the best model is selected for the test step based on the validation set loss. According to the plot, the best model is at epoch 80.

### III. EXPERIMENTAL SETUPS AND RESULT

The ISBI 2015 dataset evaluates the proposed network and a direct comparison is made with plenty of state-of-the-art approaches. The main reason for choosing this dataset is the challenge posed by lesions, in terms of size, shape, and location. Therefore, future sections shall discuss evaluation criteria and outcomes.

#### A. Evaluation Metrics

Generally, with the usage of the model, the metrics between the ground truth and the predicted image measure the evaluation metrics of segmentation methods. Hence, the following describes the evaluation metrics employed by the current research. With the assumption that $\mathcal{M}_A$ represents the ground truth created by one of the experts and $\mathcal{M}_R$ is the output generated by the model, each of the evaluation metrics is defined as [46]:

**Dice Similarity Coefficient (DSC):** Dice is a commonly used volume metric for measuring the similarity between the ground truth and the output of the model.

$$DSC(\mathcal{M}_R, \mathcal{M}_A) = \frac{2|\mathcal{M}_A \cap \mathcal{M}_R|}{|\mathcal{M}_A| + |\mathcal{M}_R|} \quad (4)$$

**Lesion True Positive Rate (LTPR):** With the assumption of the list of lesions, $\mathcal{L}_R$, as the 18-connected components of $\mathcal{M}_R$ and a similar definition for $\mathcal{L}_A$, the lesion-wise ratio of true positives, LTPR, is defined as:

$$LTPR(\mathcal{M}_R, \mathcal{M}_A) = \frac{|\mathcal{L}_R \cap \mathcal{L}_A|}{|\mathcal{L}_R \cap \mathcal{L}_A| + |\mathcal{L}_R \cap \mathcal{L}_A^C|} \quad (5)$$

Where $\mathcal{L}_A^C$ is the complement of $\mathcal{L}_A$.

**Lesion False Positive Rate (LFPR):** LFPR is the lesion-wise ratio of false positives which is given as:

$$LFPR(\mathcal{M}_R, \mathcal{M}_A) = \frac{|\mathcal{L}_R^C \cap \mathcal{L}_A|}{|\mathcal{L}_R^C \cap \mathcal{L}_A| + |\mathcal{L}_R^C \cap \mathcal{L}_A|} \quad (6)$$

**Absolute Volume Difference (AVD):** The total volume of the absolute difference is divided by the total volume of the ground truth.

$$AVD(\mathcal{M}_R, \mathcal{M}_A) = \frac{\max(|\mathcal{M}_R|, |\mathcal{M}_A|) - \min(|\mathcal{M}_R|, |\mathcal{M}_A|)}{\mathcal{M}_R} \quad (7)$$

Although providing more information (anatomical and tissue-based features) for learning-based methods can lead to getting effective and accurate learning, the goal of this paper is not only to achieve high accuracy but also to provide a system with a minimum of modality and at the same time high accuracy. Therefore, the number of input modalities acts as one of the evaluation metrics, because, as the number of input modalities lowers, the method becomes more beneficial for patients in terms of cost, time, and usability.

#### B. Results

The proposed method's efficiency on the ISBI dataset is evaluated by a process that is carried out in two stages.

In the first stage, the evaluation is with training data, in which the ground truth of the images is available. Table 1 provides the results of the comparison to those of other methods. As seen in the table, the current study's approach outperforms other methods in terms of DSC and LTPR. For images of high and low lesion loads, Figure 8 compares the present paper's segmentation results to ground truths.

TABLE 1: THE FIRST ISBI DATA RESULTS OF THE PROPOSED MODEL IN COMPARISON TO THOSE OF OTHER MODELS. IN THIS EXPERIMENT, THE ISBI DATASET INCLUDES IMAGES WITH AVAILABLE GROUND TRUTH. THE MEAN VALUES OF DSC, LTPR, AND LFPR FOR DIFFERENT METHODS ARE SHOWN.

| Method | R1 | | | R2 | | |
|---|---|---|---|---|---|---|
| | Dice | LTPR | LFPR | Dice | LTPR | LFPR |
| R1 | - | - | - | 0.7320 | 0.6550 | 0.1740 |
| R2 | 0.7320 | 0.8260 | 0.3550 | - | - | - |
| Brosch et al. [29] (Trained by R1) | 0.6844 | 0.7455 | 0.5455 | 0.6444 | 0.6333 | 0.5288 |
| Brosch et al. [29] (Trained by R2) | 0.6833 | 0.7833 | 0.6455 | 0.6588 | 0.6933 | 0.6199 |
| Aslani et al.[45] (Trained by R1) | 0.6980 | 0.7460 | 0.4820 | 0.6510 | 0.6410 | 0.4506 |
| Aslani et al.[45] (Trained by R2) | 0.6940 | 07840 | 0.4970 | 0.6640 | 0.6950 | 0.4420 |
| Aslani et al.[30] (Trained by R1) | 0.7649 | 0.6697 | **0.1202** | 0.6989 | 0.5356 | **0.1227** |
| Aslani et al.[30] (Trained by R2) | 0.7646 | 0.7002 | 0.2022 | 0.7128 | 0.5723 | 0.1896 |
| Proposed-Method (One Modality) | **0.7982** | 0.8013 | 0.3676 | **0.7978** | 0.7295 | 0.2628 |
| Proposed-Method (Two Modalities) | 0.7865 | **0.8017** | 0.3923 | 0.7856 | **0.7298** | 0.2945 |

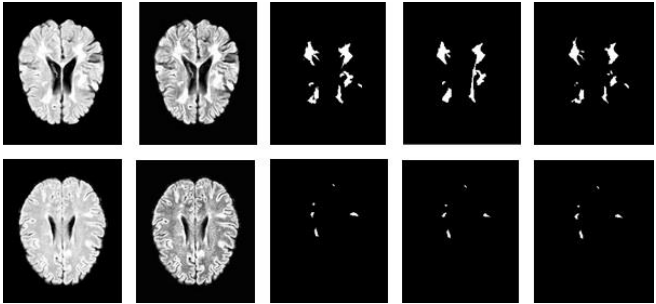

Figure 8: Output segmentation by the present study on two ISBI training dataset subjects (the first row is Subject 2 and the second row is Subject 3) compared to the ground truth provided by Rater 1 and Rater 2. From left to right, the first column is the original FLAIR image, the second column is after the preprocessing method, the third column is the segmentation result of the proposed method, and the next two columns are the ground truth provided by Rater 1 and Rater 2.

In the second stage, ground truths are not available in the ISBI test set (with 14 subjects) used to evaluate the proposed method and the evaluation metrics are calculated utilizing the challenge web service. The current work trains its model with the training data's four subjects, one of which serves as the validation data. Then, the segmentation of the test set is predicted and, finally, the 3D segmentation results are submitted to the challenge web service for evaluation. Table 2 presents the results of the ISBI test set and compares these to those of other published papers. Clearly, in some of the evaluation metrics, the present study's results for the two modalities are superior to those of other studies. Even in the single modality (the FLAIR image), the current paper's results are satisfactory when compared to the findings of other approaches.

As mentioned in Section 3, there are two different ground truths for each training sample and this difference indicates the challenge of accurately labeling lesion areas. In the proposed method, the knowledge of both experts is employed to train the model. As seen in Table 2, in comparison to other studies, the proposed method has a high LFPR. However, the visualization of the results shows that most FP pixels are in the connected neighborhood of TP pixels. In other words, the algorithm is unlikely to predict non-lesion pixels as lesions unless these pixels are connected to a lesion area, thus producing a slight increase in the lesion area. As shown in Figure 9, the created false positive pixels are all connected in the vicinity of real positive pixels, which slightly expands the area of the lesion.

TABLE 2: THE RESULTS OF THE PROPOSED METHOD ON THE OFFICIAL ISBI TEST SET WHEN COMPARED TO THE RESULTS OF OTHER METHODS. THE METRICS WITH THE BEST AND SECOND-BEST PERFORMANCES ARE INDICATED BY BOLD AND UNDERLINED VALUES, RESPECTIVELY.

| Method | Modalities | Dice | LTPR | LFPR | AVD |
|---|---|---|---|---|---|
| Andermatt et al. [42] | 4 (T1, T2, FLAIR, PD) | 0.6298 | **0.4871** | 0.2013 | 0.4045 |
| Hu et al. [31] | 4 (T1, T2, FLAIR, PD) | <u>0.6345</u> | <u>0.4787</u> | 0.1299 | 0.3949 |
| Birenbaum et al. [37] | 4 (T1, T2, FLAIR, PD) | 0.6271 | 0.5678 | 0.4975 | 0.3585 |
| Hashemi et al.[44] | 4 (T1, T2, FLAIR, PD) | 0.5841 | 0.4558 | **0.0866** | 0.4972 |
| Valverde et al.[24] | 3 (T1, T2, FLAIR) | 0.6304 | 0.3669 | 0.1529 | **0.3384** |
| Aslani et al.[30] | 3 (T1, T2, FLAIR) | 0.6114 | 0.4103 | 0.1393 | 0.4537 |
| Salem et al.[43] | 3(T1, T2, FLAIR) | 0.6305 | 0.3670 | 0.1529 | 0.3585 |
| Roy et al. [32] | 2 (T1, FLAIR) | 0.5243 | ---- | <u>0.1103</u> | 0.5207 |
| Valverde et al. [41] | 2 (T1, FLAIR) | 0.5774 | 0.2967 | 0.1885 | 0.3848 |
| Proposed Method | 2 (T1, FLAIR) | **0.6430** | 0.4543 | 0.3524 | <u>0.3524</u> |
| Proposed Method | **1 (FLAIR)** | 0.6321 | 0.4547 | 0.3868 | 0.3880 |

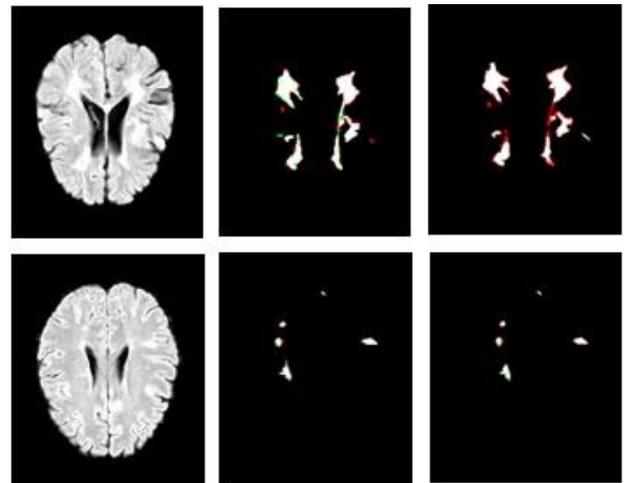

Figure 9: Output segmentation of the introduced model in comparison to two ground truths. Left to right are the FLAIR image, the results compared to the first expert, and the results compared to the second expert. In all images, true positives are denoted in white pixels, false positives in red pixels, and false negatives in green pixels.

The boxplots of the DSC, LFPR, LTPR and AVD evaluation metrics for different approaches are illustrated in Fig 10. The Figure shows our proposed method performs well in terms of DSC and AVD compare to other state-of-the-art methods. As it is obvious, the DSC is in the range of [0.3, 0.85] and most of the samples are over 0.6. In addition, although the mean of the LFPR is high for the proposed method, there are no samples in which the LFPR is over 0.7, and most of the samples are located under 0.45.

Figure 11 shows the relationship between the ground-truth and predicted volumes of lesions (each point per lesion). In terms of qualitative evaluation, it can be seen that most of the methods tend to underestimate lesions as many of the points are under the red line especially when the size of lesions is increasing. On the contrary, the method which is proposed by Roy [32] tends to overestimate the lesion sizes. However, with quantitative analysis, the proposed method estimated lesion size the slope closest to unity (0.76) together with the highest Pearson correlation coefficient (0.94). So, it means the proposed method provides a stronger global agreement between predicted lesion volumes and ground-truth lesion volumes. It is worth mentioning, a better agreement does not mean the model has better accuracy.

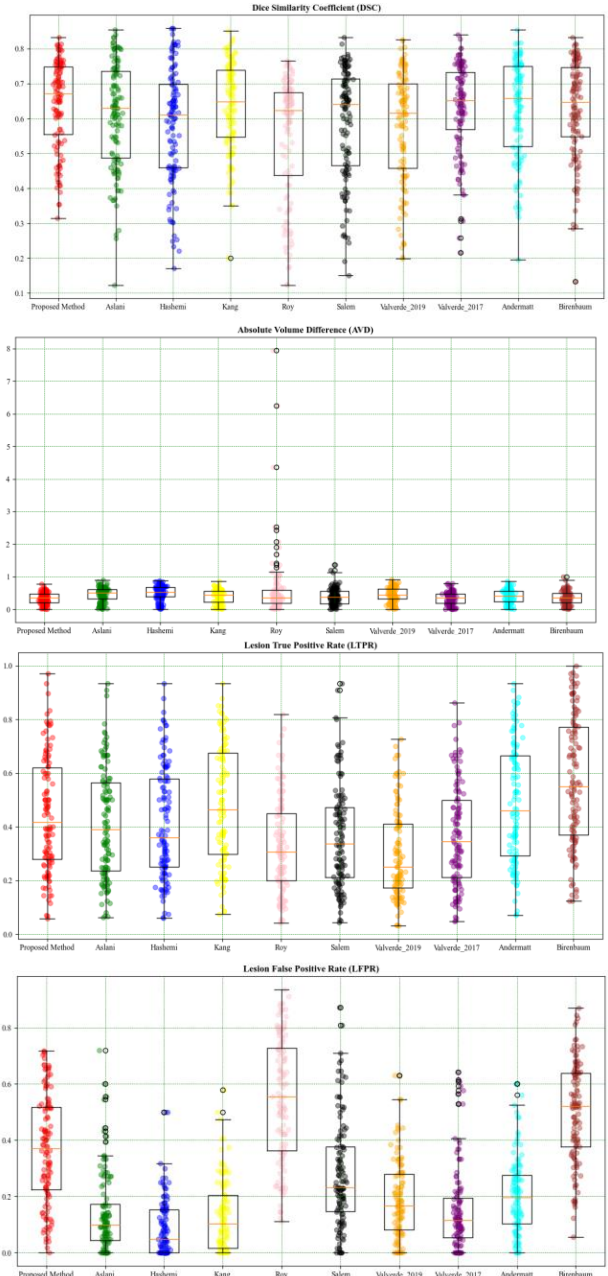

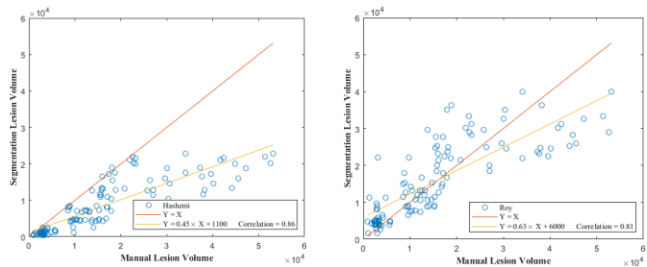

Figure 10: Demonstration of the tested models' boxplot with all measures on the ISBI dataset.

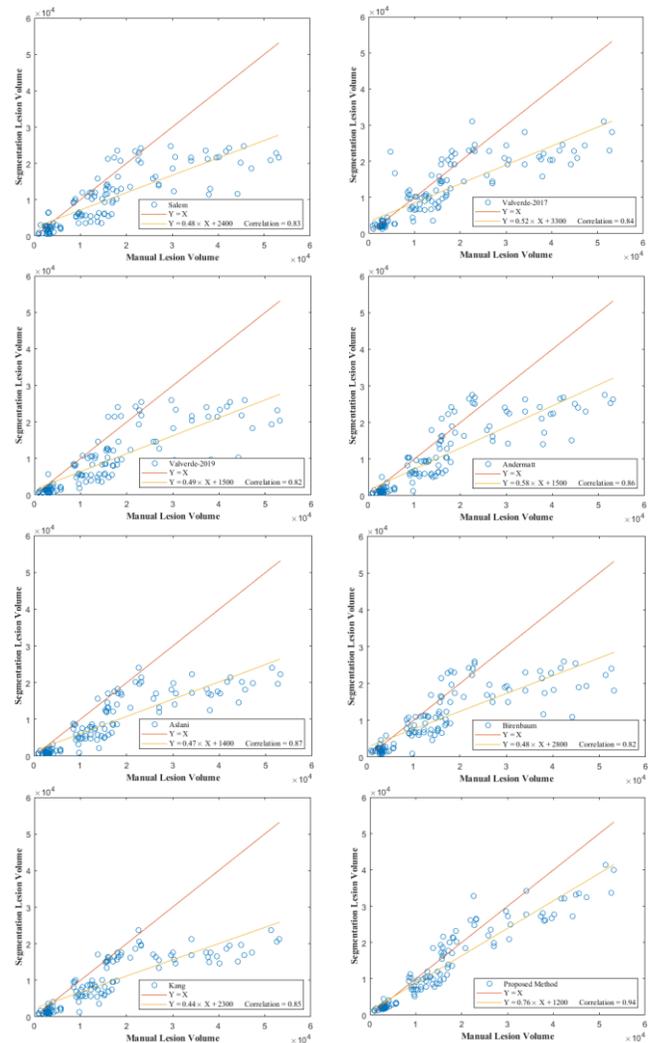

Figure 11: Comparing the lesion volumes produced by automatic and manual segmentation on the ISBI dataset. Each point is related to a single lesion.

## IV. DISCUSSION AND CONCLUSIONS

The present paper introduces an automated lesion segmentation approach that is based on one modality as an input which is rare in the state-of-the-art. The proposed architecture is an end-to-end 3D patch-wise composed of convolution, deconvolution, and an SCA-VoxRes module as an attention module.

In medical image processing, it is common to overcome the limitation of the single modality approach by combining different MRI modalities. Holding that patients cannot afford the cost of combining modalities in medical image segmentation, the current research presents a single modality-based architecture that is more accurate and robust than other multi-modality methods.

Furthermore, patch size is the limitation of patch-based CNNs which suffer from a lack of spatial information for the lesion. Therefore, the proposed method uses an attention module to capture long-range contextual information as a way of attaining spatial information. Consequently, the patch-based CNNs do not lack data, because many random patches can be extracted from 3D data.

Although there are several advantages to the proposed method the limitation of the proposed method should be taken into account. The most important limitation of the proposed method is the high range of LFPR which is due to the existence of two different grand truths in dataset. In this case, when the model is training, it should try to optimize the loss function according to the logical-or of two grand truths.

REFERENCES

[1] L. Steinman, Multiple sclerosis: a coordinated immunological attack against myelin in the central nervous system, Cell 85 (3) (1996) 299–302. doi:10.1016/s0092-8674(00)81107-1.

[2] A. Compston, H. Winedl, B. Kieseier, Coles, Multiple sclerosis. Lancet 372 (2008) 1502–1517. doi:10.1016/S0140-6736(02)08220-X.

[3] L. A. Rolak, Multiple sclerosis: it's not the disease you thought it was, Clinical Medicine & Research 1 (1) (2003) 57–60. doi:10.3121/cmr.1.1.57.

[4] J. Simon, D. Li, A. Traboulsee, P. Coyle, D. Arnold, F. Barkhof, J. Frank, R. Grossman, D. Paty, E. Radue, et al., Standardized mr imagingprotocol for multiple sclerosis: Consortium of ms centers consensus guidelines, American Journal of Neuroradiology 27 (2) (2006) 455–461.

[5] E. M. Sweeney, R. T. Shinohara, N. Shiee, F. J. Mateen, A. A. Chudgar, J. L. Cuzzocreo, P. A. Calabresi, D. L. Pham, D. S. Reich, C. M. Crainiceanu, Oasis is automated statistical inference for segmentation, with applications to multiple sclerosis lesion segmentation in mri, NeuroImage: clinical 2 (2013) 402–413. doi:10.1016/j.nicl.2013.03.002.

[6] ] M. Cabezas, A. Oliver, S. Valverde, B. Beltran, J. Freixenet, J. C. Vilanova, L. Ramió-Torrentà, À. Rovira, X. Lladó, Boost: A supervised approach for multiple sclerosis lesion segmentation, Journal of neuroscience methods 237 (2014) 108–117. doi:10.1016/j.jneumeth.2014.08.024.

[7] E. Geremia, O. Clatz, B. H. Menze, E. Konukoglu, A. Criminisi, N. Ayache, Spatial decision forests for ms lesion segmentation in multi-channel magnetic resonance images, NeuroImage 57 (2) (2011) 378–390. doi:10.1016/j.neuroimage.2011.03.080.

[8] A. Jesson, T. Arbel, Hierarchical mrf and random forest segmentation of ms lesions and healthy tissues in brain mri, Proceedings of the 2015 longitudinal multiple sclerosis lesion segmentation challenge (2015) 1–2.

[9] X. Lladó, A. Oliver, M. Cabezas, J. Freixenet, J. C. Vilanova, A. Quiles, L. Valls, L. Ramió-Torrentà, À. Rovira, Segmentation of multiple sclerosis lesions in brain mri: a review of automated approaches, Information Sciences 186 (1) (2012) 164–185. doi:10.1016/j.ins.2011.10.011.

[10] N. Guizard, P. Coupé, V. S. Fonov, J. V. Manjón, D. L. Arnold, D. L. Collins, Rotation-invariant multi-contrast non-local means for ms lesion segmentation, NeuroImage: Clinical 8 (2015) 376–389. doi:10.1016/j.nicl.2015.05.001.

[11] ] M. D. Steenwijk, P. J. Pouwels, M. Daams, J. W. van Dalen, M. W. Caan, E. Richard, F. Barkhof, H. Vrenken, Accurate white matter lesion segmentation by k nearest neighbor classification with tissue type priors (knn-ttps), NeuroImage: Clinical 3 (2013) 462–469. doi:10.1016/j.nicl.2013.10.003.

[12] M. J. Fartaria, G. Bonnier, A. Roche, T. Kober, R. Meuli, D. Rotzinger, R. Frackowiak, M. Schluep, R. Du Pasquier, J.-P. Thiran, et al., Automated detection of white matter and cortical lesions in early stages of multiple sclerosis, Journal of Magnetic Resonance Imaging 43 (6) (2016) 1445–1454. doi:10.1002/jmri.25095.

[13] X. Tomas-Fernandez, S. K. Warfield, A model of population and subject (mops) intensities with application to multiple sclerosis lesion segmentation, IEEE transactions on medical imaging 34 (6) (2015) 1349–1361. doi:10.1109/TMI.2015.2393853.

[14] P. Schmidt, C. Gaser, M. Arsic, D. Buck, A. Förschler, A. Berthele, M. Hoshi, R. Ilg, V. J. Schmid, C. Zimmer, et al., An automated tool for detection of flair-hyperintense white-matter lesions in multiple sclerosis, Neuroimage 59 (4) (2012) 3774–3783. doi:10.1016/j.neuroimage.2011.11.032.

[15] E. Roura, A. Oliver, M. Cabezas, S. Valverde, D. Pareto, J. C. Vilanova, L. Ramió-Torrentà, À. Rovira, X. Lladó, A toolbox for multiple sclerosis lesion segmentation, Neuroradiology 57 (10) (2015) 1031–1043. doi:10.1007/s00234-015-1552-2.

[16] R. Harmouche, N. K. Subbanna, D. L. Collins, D. L. Arnold, T. Arbel, Probabilistic multiple sclerosis lesion classification based on modeling regional intensity variability and local neighborhood information, IEEE transactions on Biomedical Engineering 62 (5) (2014) 1281–1292. doi:10.1109/TBME.2014.2385635.

[17] M. Strumia, F. R. Schmidt, C. Anastasopoulos, C. Granziera, G. Krueger, T. Brox, White matter ms-lesion segmentation using a geometric brain model, IEEE transactions on medical imaging 35 (7) (2016) 1636–1646. doi:10.1109/TMI.2016.2522178

[18] Y. LeCun, L. Bottou, Y. Bengio, P. Haffner, Gradient-based learning applied to document recognition, Proceedings of the IEEE 86 (11) (1998) 2278–2324. doi:10.1109/5.726791

[19] Y. LeCun, Y. Bengio, G. Hinton, Deep learning. nature (2015), May; 521 (7553): 436 10.1038/nature14539doi:10.1038/nature14539.

[20] K. Simonyan, A. Zisserman, Very deep convolutional networks for large-scale image recognition, arXiv preprint arXiv:1409.1556.

[21] A. Krizhevsky, I. Sutskever, G. E. Hinton, Imagenet classification with deep convolutional neural networks, Advances in neural information processing systems 25 (2012) 1097–1105. doi:10.1145/3065386.

[22] X.-H. Han, J. Lei, Y.-W. Chen, Hep-2 cell classification using k-support spatial pooling in deep cnns, in: Deep Learning and Data Labeling for Medical Applications, Springer, 2016, pp. 3–11. doi:10.1007/978-3-319-46976-8_1.

[23] P. Liskowski, K. Krawiec, Segmenting retinal blood vessels with deep neural networks, IEEE transactions on medical imaging 35 (11) (2016) 2369–2380. doi:10.1109/TMI.2016.2546227.

[24] S. Valverde, M. Cabezas, E. Roura, S. González-Villà, D. Pareto, J. C. Vilanova, L. Ramió-Torrentà, À. Rovira, A. Oliver, X. Lladó, Improving automated multiple sclerosis lesion segmentation with a cascaded 3d convolutional neural network approach, NeuroImage 155 (2017) 159–168. doi:10.1016/j.neuroimage.2017.04.034.

[25] M. Havaei, A. Davy, D. Warde-Farley, A. Biard, A. Courville, Y. Bengio, C. Pal, P.-M. Jodoin, H. Larochelle, Brain tumor segmentation with deep neural networks, Medical image analysis 35 (2017) 18–31. doi:10.1016/j.media.2016.05.004.

[26] O. Ronneberger, P. Fischer, T. Brox, U-net: Convolutional networks for biomedical image segmentation, in: International Conference on Medical image computing and computer-assisted intervention, Springer, 2015, pp. 234–241. doi:10.1007/978-3-319-24574-4_28.

[27] P. Moeskops, M. A. Viergever, A. M. Mendrik, L. S. De Vries, M. J. Benders, I. Išgum, Automatic segmentation of mr brain images with a convolutional neural network, IEEE transactions on medical imaging 35 (5) (2016) 1252–1261. doi:10.1109/TMI.2016.2548501.

[28] K.-L. Tseng, Y.-L. Lin, W. Hsu, C.-Y. Huang, Joint sequence learning and cross-modality convolution for 3d biomedical segmentation, in: Proceedings of the IEEE conference on Computer Vision and Pattern Recognition, 2017, pp. 6393–6400. doi:10.1109/CVPR.2017.398.

[29] T. Brosch, L. Y. Tang, Y. Yoo, D. K. Li, A. Traboulsee, R. Tam, Deep 3d convolutional encoder networks with shortcuts for multiscale feature integration applied to multiple sclerosis lesion segmentation, IEEE transactions on medical imaging 35 (5) (2016) 1229–1239. doi:10.1109/TMI.2016.2528821.

[30] S. Aslani, M. Dayan, L. Storelli, M. Filippi, V. Murino, M. A. Rocca, D. Sona, Multi-branch convolutional neural network for multiple sclerosis lesion segmentation, NeuroImage 196 (2019) 1–15. doi:10.1016/j.neuroimage.2019.03.068.

[31] G. Kang, B. Hou, Y. Ma, F. Labeau, Z. Su, et al., Acu-net: A 3d attention context u-net for multiple sclerosis lesion segmentation, in: ICASSP 2020-2020 IEEE International Conference on Acoustics, Speech and Signal Processing (ICASSP), IEEE, 2020, pp. 1384–1388. doi:10.1109/ICASSP40776.2020.9054616.

[32] S. Roy, J. A. Butman, D. S. Reich, P. A. Calabresi, D. L. Pham, Multiple sclerosis lesion segmentation from brain mri via fully convolutional neural networks, arXiv preprint arXiv:1803.09172.

[33] H. Chen, Q. Dou, L. Yu, J. Qin, P.-A. Heng, Voxresnet: Deep voxelwise residual networks for brain segmentation from 3d mr images, NeuroImage 170 (2018) 446–455. doi:10.1016/j.neuroimage.2017.04.041.

[34] J. Fu, J. Liu, H. Tian, Y. Li, Y. Bao, Z. Fang, H. Lu, Dual attention network for scene segmentation, in: Proceedings of the IEEE/CVF Conference on Computer Vision and Pattern Recognition, 2019, pp. 3146–3154. doi:10.1109/CVPR.2019.00326.

[35] J. Hu, L. Shen, G. Sun, Squeeze-and-excitation networks, in: Proceedings of the IEEE conference on computer vision and pattern recognition, 2018, pp. 7132–7141. doi:10.1109/CVPR.2018.00745.

[36] M. Ghafoorian, B. Platel, Convolutional neural networks for ms lesion segmentation, method description of diag team, Proceedings of the 2015 Longitudinal Multiple Sclerosis Lesion Segmentation Challenge (2015) 1–2.

[37] A. Birenbaum, H. Greenspan, Longitudinal multiple sclerosis lesion segmentation using multi-view convolutional neural networks, in: Deep Learning and Data Labeling for Medical Applications, Springer, 2016, pp. 58–67. doi:10.1007/978-3-319-46976-8_7.

[38] H. Afzal, S. Luo, S. Ramadan, J. Lechner-Scott, M. R. Amin, J. Li, M. K. Afzal, Automatic and robust segmentation of multiple sclerosis lesions with convolutional neural networks, CMC-COMPUTERS MATERIALS & CONTINUA 66 (1) (2021) 977–991. doi:10.32604/cmc.2020.012448.

[39] Y. Shachor, H. Greenspan, J. Goldberger, A mixture of views network with applications to multi-view medical imaging, Neurocomputing 374 (2020) 1–9. doi:10.1016/j.neucom.2019.09.027.

[40] S. Vaidya, A. Chunduru, R. Muthuganapathy, G. Krishnamurthi, Longitudinal multiple sclerosis lesion segmentation using 3d convolutional neural networks, Proceedings of the 2015 Longitudinal Multiple Sclerosis Lesion Segmentation Challenge (2015) 1–2.

[41] S. Valverde, M. Salem, M. Cabezas, D. Pareto, J. C. Vilanova, L. Ramió-Torrentà, À. Rovira, J. Salvi, A. Oliver, X. Lladó, One-shot domain adaptation in multiple sclerosis lesion segmentation using convolutional neural networks, NeuroImage: Clinical 21 (2019) 101638. doi:10.1016/j.nicl.2018.101638.

[42] S. Andermatt, S. Pezold, P. C. Cattin, Automated segmentation of multiple sclerosis lesions using multi-dimensional gated recurrent units, in: International MICCAI Brainlesion Workshop, Springer, 2017, pp. 31–42. doi:10.1007/978-3-319-75238-9_3.

[43] M. Salem, S. Valverde, M. Cabezas, D. Pareto, A. Oliver, J. Salvi, À. Rovira, X. Lladó, Multiple sclerosis lesion synthesis in mri using an encoder-decoder u-net, IEEE Access 7 (2019) 25171–25184. doi:10.1109/ACCESS.2019.2900198.

[44] S. R. Hashemi, S. S. M. Salehi, D. Erdogmus, S. P. Prabhu, S. K. Warfield, A. Gholipour, Asymmetric loss functions and deep denselyconnected networks for highly-imbalanced medical image segmentation: Application to multiple sclerosis lesion detection, IEEE Access 7 (2018) 1721–1735. doi:10.1109/ACCESS.2018.2886371.

[45] S. Aslani, M. Dayan, V. Murino, D. Sona, Deep 2d encoder-decoder convolutional neural network for multiple sclerosis lesion segmentation in brain mri, in: International MICCAI Brainlesion Workshop, Springer, 2018, pp. 132–141. doi:10.1007/978-3-030-11723-8_13.

[46] A. Carass, S. Roy, A. Jog, J. L. Cuzzocreo, E. Magrath, A. Gherman, J. Button, J. Nguyen, F. Prados, C. H. Sudre, et al., Longitudinal multiple sclerosis lesion segmentation: resource and challenge, NeuroImage 148 (2017) 77–102. doi:10.1016/j.neuroimage.2016.12.064.

[47] M. Jenkinson, P. Bannister, M. Brady, S. Smith, Improved optimization for the robust and accurate linear registration and motion correction of brain images, Neuroimage 17 (2) (2002) 825–841. doi:10.1016/S1053-8119(02)91132-8.

[48] J. G. Sled, A. P. Zijdenbos, A. C. Evans, A nonparametric method for automatic correction of intensity nonuniformity in mri data, IEEE transactions on medical imaging 17 (1) (1998) 87–97.

[49] K. Oishi, K. Zilles, K. Amunts, A. Faria, H. Jiang, X. Li, K. Akhter, K. Hua, R. Woods, A. W. Toga, et al., Human brain white matter atlas: identification and assignment of common anatomical structures in superficial white matter, Neuroimage 43 (3) (2008) 447–457. doi:10.1016/j.neuroimage.2008.07.009.

[50] S. M. Pizer, E. P. Amburn, J. D. Austin, R. Cromartie, A. Geselowitz, T. Greer, B. ter Haar Romeny, J. B. Zimmerman, K. Zuiderveld, Adaptive histogram equalization and its variations, Computer vision, graphics, and image processing 39 (3) (1987) 355–368. doi: 10.1016/S0734-189X(87)80186-X.

[51] T. Zhou, S. Ruan, S. Canu, A review: Deep learning for medical image segmentation using multi-modality fusion, Array 3 (2019) 100004. doi:10.1016/j.array.2019.100004.

[52] T.-Y. Lin, P. Goyal, R. Girshick, K. He, P. Dollár, Focal loss for dense object detection, in: Proceedings of the IEEE international conference on computer vision, 2017, pp. 2980–2988. doi:10.1109/TPAMI.2018.2858826.

[53] D. P. Kingma, J. Ba, Adam: A method for stochastic optimization, arXiv preprint arXiv:1412.6980.

[54] J. Duchi, E. Hazan, Y. Singer, Adaptive subgradient methods for online learning and stochastic optimization., Journal of machine learning research 12 (7). doi:10.5555/1953048.2021068.

[55] M. D. Zeiler, Adadelta: an adaptive learning rate method, arXiv preprint arXiv:1212.5701.

[56] Y. N. Dauphin, H. De Vries, Y. Bengio, Equilibrated adaptive learning rates for non-convex optimization, arXiv preprint arXiv:1502.04390.

**Mehdi SadeghiBakhi** received the B.S. degree in Computer Engineering from Yazd University, Iran and received his M.S. degree in artificial intelligence from Ferdowsi university of Iran, Mashhad. His research interests include Medical Image Processing, Deep Learning and Machine Learning.

**Prof. HamidReza Pourreza** is currently Professor of Computer Science and Engineering at Ferdowsi University of Mashhad (FUM). He received his B.S. degree in Electrical Engineering from FUM in 1989, and received his M.S. and Ph.D. degrees in Electrical Engineering Computer Engineering from Amirkabir University of Technology in 1993 and 2003, respectively. His research interests are in the areas of Deep Learning, Computer Vision, Hardware Design, and Intelligent Transportation Systems (ITS).

**Prof. HamidReza Mahyar** is currently Professor of Computer Science and Engineering at McMaster. He received his Ph.D. degree in Computer Engineering from Sharif University of Technology (SUT). His research interests lie at the intersection of machine learning, network science, and natural language processing, with a current emphasis on the fast-growing subject of graph neural networks with applications in social networks, recommendation systems, and drug discovery.